\documentstyle[prl,twocolumn,aps]{revtex}
\begin{document}
\draft
\title{On the detection of condensate vortex states}
\author{E. V. Goldstein, E. M. Wright, and P. Meystre}
\address{Optical Sciences Center, University of Arizona, Tucson, AZ 85721\\
December 11, 1997 \\ \medskip}
\author{\small\parbox{14.2cm}{\small\hspace*{3mm}
We discuss a detection scheme that reveals the existence of vortex
states in a cylindrically symmetric condensate trap. It relies on the
measurement of the second-order correlation function of the Schr\"odinger
field, and yields directly the topological charge of the vortex state.
\\[3pt]PACS numbers: 03.75.-b 03.75.Fi 42.50.Vk 67.40.Vs}}
\maketitle
\narrowtext

The rapid progress in the experimental generation and manipulation of
Bose-Einstein condensates in low density trapped alkali vapors
\cite{AndEnsMat95,DavMewAnd95,EnsJinMat96,MewAndDru96,BraSacTolHul95} opens
up the way to the detailed study of the thermodynamic and dynamical
properties of weakly interacting quantum-degenerate gases. A topic of much
current interest is the superfluidity of these samples. Superfluidity is
related to the rotational properties, and in particular to the existence
of vortex states in quantum gases \cite{TilTil90}.
While no vortex states have been launched
in low density atomic condensates so far, their existence has been
numerically  studied \cite{EdwDodCla96,Str961,DalStr96} and their
stability analyzed \cite{Rok97}. Optical methods to launch vortex states
were recently proposed in Ref. \cite{MarZhaWri97,BolWal98}, and Jackson
et. al. \cite{JacMcCAda97} have produced numerical solutions of vortex
formation in a weakly-interacting condensate by piercing it
and subsequently slicing it
with a blue-detuned laser at a velocity exceeding the critical velocity.

In view of these developments, it is therefore timely and important to
discuss ways to detect vortices in trapped atomic condensates.
One possibility suggested in Ref. \cite{MarZhaWri97} is based on
measuring the spatial absorption profile of the sample. This method is
sensitive to the rotation of the condensate via the rotational Doppler shift
\cite{BiaBia97}, but does not give a direct measure of its topological charge.
Rather, the quantization of the vortex motion must be inferred from other
condensate parameters such as number of particles, system size, etc.
In contrast, the scheme that we propose and analyze in this
note produces a spatial interference pattern that results directly from
the nonzero atomic angular momentum of the vortex and gives a direct
measurement of its topological charge. In addition, a slight variation
of that method allows a direct demonstration of the existence of persistent
currents in the sample.
The method is based on the ionization detection
scheme discussed in \cite{GolMey98}, whereby one measures normally
ordered correlation functions of the Schr\"odinger field operator
${\hat \Psi}({\bf r},t)$. Specifically, we show that the two-point
second-order correlation functions provide direct information on the
persistent currents and the topological charge of the vortex state.

The proposed schemes \cite{MarZhaWri97,BolWal98} to create vortices in
cylindrically symmetric traps rely on Raman transitions between two
hyperfine levels of ground state alkali atoms, the effective coupling
resulting from virtual electric dipole transitions induced by far-off
resonant laser beams. For this discussion we assume that the system is
effectively two-dimensional as a result of strong confinement in the
z-direction.  A coherent rotational coupler is realized if the
lasers used for the Raman transitions carry angular
momentum associated with their field envelope, which can be
transfered to the center-of-mass motion of the atoms.
This can be achieved e.g. \cite{MarZhaWri97} by placing the condensate
at the focus $z=0$ of two Laguerre-Gaussian
beams of polarizations $\sigma^+$ and $\sigma^-$ propagating along the $z$
direction and with Rabi frequencies
\begin{equation}
\Omega_\pm({\bf r})=\Omega_0 \left (\sqrt{2}\frac{\rho}{w} \right )^{|m|}
L_p^{|m|}\left (2\frac{\rho^2}{w^2} \right )
e^{-\rho^2/w^2} e^{\pm i m\varphi} e^{ikz}
\label{beam}
\end{equation}
where $\rho^2=x^2+y^2$, $w$ is the beam spot size, and $\varphi$ is the
azimuthal angle. The transverse structure of these fields is
characterized by the radial mode number $p$ and the azimuthal mode
number $m$ \cite{KogLi66}.  In particular, they
have a phase singularity at their center
with topological charge $\pm m$ and associated angular momentum
$\pm \hbar m$.  As a result of the Raman transitions the laser fields
couple a condensate in the ground state
$\langle {\bf r}|g \rangle = \psi_g({\bf r})|F=1,M_F=-1\rangle$ with no
angular momentum, where $\psi_g({\bf r})$ is the ground state wave function
of the trapping potential, to the vortex state
$\langle {\bf r}|v \rangle = \psi_v({\bf r}) |F=1,M_F=1\rangle$ with
$2\hbar m$ angular momentum per atom and topological charge $2m$.
For the geometry discussed in Ref. \cite{MarZhaWri97} and using $m=1$
the vortex wave function is given explicitly by
\begin{equation}
\psi_v({\bf r})={\cal N}\rho^2e^{i\theta({\bf r})}\psi_g({\bf \rho}) ,
\label{vort}
\end{equation}
with ${\cal N}$ a normalization constant, and $\theta({\bf r})=2\varphi$.
The key point from this
discussion is that the vortex state created in this manner has a
topological charge $2m$ and associated azimuthal variation
$\exp(2im\varphi)$ in the general case.

Since $ \langle v|g \rangle = 0$, this system can conveniently be described in
terms of the spinor ${\bbox \phi}({\bf r}, t)=col(\phi_g({\bf r}, t),
\phi_v({\bf r}, t))$.  Then in the mean-field Hartree approximation, and
assuming that many-body effects do not significantly alter the spatial
profiles of the ground state and vortex state so that we may restrict
ourselves to these two modes, the components of the spinor evolve according
to
\begin{equation}
\phi_v({\bf r},t)=\beta(t) \psi_v({\bf r})  ,
\qquad
\phi_g({\bf r},t)=\alpha(t)\psi_g({\bf r})  .
\label{td}
\end{equation}
For the case $m=1$ $\alpha(t)$ and $\beta(t)$ are governed by the
equations of motion (5) and (6) of Ref. \cite{MarZhaWri97}, and this
serves as a concrete example.

More generally, in the following we consider a generic situation where the
external perturbation responsible for the coupling creates a quantum 
degenerate gas in the superposition (\ref{td}), but with the phase 
$\theta({\bf r})$ of the excited state in Eq. (\ref{vort}) kept 
general.  This will allow us to point out the specific signatures of a vortex state 
that distinguish it from a non-vortex state.

In the two-mode approximation the condensate is described by the
two-component Schr\"odinger field operator ${\hat {\bbox \Psi}}({\bf r})
= ( {\hat \Psi}_g({\bf r}), {\hat \Psi}_v({\bf r}))$ satisfying the boson
commutation relation
$[{\hat \Psi}_\ell({\bf r}),\Psi^\dagger_{\ell'}({\bf r}')]=
\delta_{\ell \ell'} \delta({\bf r}-{\bf r}'), \ell = \{v, g\}$.  Assuming
then a system composed of $N$-atoms, the quantum state of the system becomes
\begin{equation}
|\Phi(t)\rangle = \frac{1}{\sqrt{N!}} \left [ \int d^3r
(\phi_g({\bf r},t){\hat \Psi}_g^\dagger({\bf r})+\phi_v({\bf r},t)
{\hat \Psi}^\dagger_v({\bf r}))
\right ]^N |0\rangle ,
\end{equation}
or, in terms of the mode creation and annihilation operators
\begin{equation}
a_\ell^\dagger = \int d^3 r \psi_\ell({\bf r}){\hat \Psi}^\dagger({\bf r}),
\qquad
a_\ell = \int d^3 r \psi_\ell^\star({\bf r}){\hat \Psi}({\bf r}),
\label{ann}
\end{equation}
we obtain
\begin{equation}
|\Phi(t)\rangle=\frac{1}{\sqrt{N!}}
\sum_k C_N^k [\alpha(t)]^k[\beta(t)]^{N-k}
(a_g^\dagger)^k(a_v^\dagger)^{N-k}|0\rangle.
\label{binom}
\end{equation}
Hence, the sample is in an entangled superposition of the
ground and vortex states, a result of the fact that the total number of
particles is conserved, but not the individual particle numbers in the two
states. This is similar to the situation of split condensates discussed in
Refs. \cite{JavWil97,MilCorWri97}.

Due to the assumed cylindrical symmetry of the trapping potential,
the existence of
a vortex state cannot be demonstrated by off-resonance
imaging,\footnote{The situation would be different if the ground and vortex states
corresponded to the same electronic state.} which measures
correlation functions of the sample density ${\hat \rho}({\bf r},t) \equiv
{\hat \Psi}^\dagger({\bf r}, t) {\hat \Psi}({\bf r}, t)$.  Since this is
simply the sum of the condensate and vortex density profiles, and
the vortex density profile is cylindrically symmetric, the density
profile does not reveal the phase singularity associated with the vortex.
What is needed instead is a measurement scheme which involves correlation
functions of ${\hat \Psi}({\bf r}, t)$ itself. As discussed in Ref.
\cite{GolMey98}, these functions can be extracted in an ionization scheme
whereby one or more tightly focused lasers are used to selectively ionize
atoms in small regions of the condensate plus vortex system.
The measurement proceeds then by
detecting the ionized atoms, which play the role of a detector field.

We consider specifically a two-point detection scheme in which two
ionizing laser beams are focussed at locations ${\bf r}_1= (\rho_1,
\varphi_1)$ and ${\bf r}_2= (\rho_2, \varphi_2)$.
For that geometry, and assuming that the lasers are focussed onto spots
small compared to the dimensions of the condensate, the ionization scheme
measures the probability $w_2$ of jointly ionizing an atom at ${\bf r}_1$
and the other at ${\bf r}_2$ as a function of normally ordered
Schr\"odinger field correlation functions whose explicit form is
\begin{eqnarray}
w_2(t,\Delta t) &\simeq& \eta({\bf r}_1,{\bf r}_2)\eta({\bf r}_2,{\bf r}_1)
\int_t^{t+\Delta t}dt_1\int_t^{t+\Delta t}dt_2
\nonumber\\
& &\langle {\hat{\bbox \Psi}}^\dagger({\bf r}_1, t_1)
{\hat {\bbox \Psi}}^\dagger
({\bf r}_2, t_2){\hat {\bbox \Psi}}({\bf r}_2, t_1)
{\hat {\bbox \Psi}}({\bf r}_1, t_2) \rangle
\nonumber \\
&+& \eta({\bf r}_1)\eta({\bf r}_2)
\int_t^{t+\Delta t}dt_1\int_t^{t+\Delta t}dt_2
\nonumber\\
& &\langle {\hat {\bbox \Psi}}^\dagger({\bf r}_1, t_1)
{\hat {\bbox \Psi}}^\dagger({\bf r}_2, t_2)
{\hat {\bbox \Psi}}({\bf r}_2, t_2)
{\hat {\bbox \Psi}}({\bf r}_1, t_1) \rangle ,
\label{w2}
\end{eqnarray}
where $\eta({\bf r})$ is the detector efficiency and $\eta({\bf r}_1,
{\bf r}_2)$ its cross-efficiency, as discussed in Ref. \cite{GolMey98}.
The first term in $w_2$ is an exchange contribution, while the second
term is the direct contribution familiar from photodetection
theory.

Equation (\ref{w2}) can be considerably simplified 
for measurement intervals $\Delta t$ small compared to the
characteristic evolution time of the condensate. In this case
we can set $t_1=t_2=t$ and  we have simply
$$
w_2 (t)= 
$$
\begin{equation}
(\Delta t)^2 [\eta({\bf r}_1)\eta({\bf r}_2)+
\eta({\bf r}_1,{\bf r}_2) \eta({\bf r}_2,{\bf r}_1)]
G^{(2)}({\bf r}_1,{\bf r}_2;t),
\label{w22}
\end{equation}
where
$$
G^{(2)}({\bf r}_1,{\bf r}_2;t)\equiv
$$
\begin{eqnarray}
&&\langle \Phi(0)|\bbox{{\hat \Psi}}^\dagger({\bf r}_1,t)
\bbox{{\hat \Psi}}^\dagger({\bf r}_2,t) 
\bbox{{\hat \Psi}}({\bf r}_2,t)
\bbox{{\hat \Psi}}({\bf r}_1,t)|\Phi(0) \rangle  
\nonumber\\
& &=\langle \Phi(t)|\bbox{{\hat \Psi}}^\dagger({\bf r}_1)
\bbox{{\hat \Psi}}^\dagger({\bf r}_2)
\bbox{{\hat \Psi}}({\bf r}_2)
\bbox{{\hat \Psi}}({\bf r}_1)|\Phi(t) \rangle.
\label{corr2}
\end{eqnarray}
As follows from the definitions of the detector self- 
and cross-efficiencies $\eta({\bf r})$ and $\eta({\bf r}_1,{\bf r}_2)$ given
in Ref. \cite{GolMey98} the term in square brackets in Eq. (\ref{w22})
does not vary azimuthally, which is important for 
the following considerations.
Specializing in anticipation of our subsequent discussion to the case
$\rho_1 = \rho_2 = \rho$ and with Eq. (\ref{ann}) this gives readily
\begin{eqnarray}
&&G^{(2)}({\bf r}_1,{\bf r}_2;t)=  
2N(N-1)|\phi_g(\rho)|^2|\phi_v(\rho)|^2|\alpha(t)|^2|\beta(t)|^2
\nonumber\\
& &\times(1+\cos[\theta(\bf{r}_2)-\theta(\bf{r}_1)]).
\label{g2}
\end{eqnarray}
The key feature
for the present discussion is the spatial dependence contained in
the explicit phase difference $[\theta({\bf r}_2) - \theta({\bf r}_1)]$.
Provided that measurements are performed at a fixed time $t$, this
dependence allows us to determine the existence of vortex motion,
in a particularly simple way, as we show below.\\

\noindent
{\em Persistent currents}\\

The hydrodynamic formulation of superfluidity \cite{LifPit89} introduces
the velocity ${\bf v}_s = (\hbar/M) {\bbox \nabla} \theta({\bf r})$,
where $M$ is the atomic mass and $\theta({\bf r})$ is the phase of the
superfluid component, in our case the vortex phase measured relative to
the ground state phase which was tacitly taken equal to zero. Vortex
states are characterized by the fact that the circulation of ${\bf v}_s$
is quantized,
\begin{equation}
\oint {\bf v}_s \cdot d{\bf l} = 2\pi n (\hbar/M) ,
\label{circ}
\end{equation}
where $n$ is an integer. In order to detect the circulation of the
vortex, it is sufficient to determine its tangential velocity
component $v_\varphi$. Hence the detectors can remain on a circle centered
on the axis of rotation of the vortex and we have
\begin{equation}
{\bf \nabla} \theta = \frac{1}{\rho} \frac{\partial \theta({\bf r})}{\partial
\varphi} {\hat {\bf e}}_\varphi  ,
\end{equation}
where ${\hat {\bf e}}_\varphi$ is the unit vector tangential to the
radial direction. For small distances $|{\bf r}_2-{\bf r}_1|$, the
relative phase appearing in the last term in the correlation
function (\ref{g2}) becomes
\begin{equation}
\theta({\bf r}_2)-\theta({\bf r}_1) \simeq
({\bf \nabla} \theta)\cdot {\hat {\bf e}}_\varphi d\rho 
= (M/\hbar) v_\varphi(\varphi) d\rho.
\end{equation}
For a general phase variation $\theta({\bf r})$ the local velocity,
and hence the current of atoms, will vary azimuthally.  However,
for a vortex of topological charge $2m$, we have
$v_\varphi=2m\hbar/M\rho$ independent of the azimuthal position
of the pair of closely spaced detectors, and $n=2m$ in Eq. (\ref{circ}).
Hence, moving the pair of detectors along a circle while keeping their
distance $d\rho$ fixed allows one to determine the presence of persistent
currents, $v_\varphi(\varphi) =$ constant.  Detection of this persistent
current is a key signature of a vortex state.  From the value of
this persistent current one could infer the value of $n$ if all other
parameters were known. Next we describe a second measurement which yields
the topological charge more directly.\\

\noindent
{\em Topological charge}\\

In addition, it is also possible to carry out a different class of
measurements where detector 1 is held at a fixed position relative to
the vortex core, while detector 2 is moved on a circle relative to
detector 1.  In that case, $G^{(2)}$ will exhibit oscillations as the
relative azimuthal angle $(\varphi_1-\varphi_2)$ between the two
detectors is varied, the phase difference being
\begin{equation}
\theta({\bf r}_1)-\theta({\bf r}_2)= 2m(\varphi_1-\varphi_2)  .
\end{equation}
Thus, the second-order correlation function
shows interference fringes as the relative azimuthal
angle of the detectors is varied, and the topological charge of the vortex
may be extracted as twice the number of (complete) bright fringes in the
interference pattern as detector 2 is moved through a full circle.
Note that in contrast to the preceding scheme, which requires knowledge
of the system parameters, e.g. the atomic mass, to determine  
$v_\varphi$ absolutely, the present method reveals the topological 
charge of the vortex, $2m$, from a global property of the interference pattern,
the number of bright fringes.

To summarize, we have discussed a measurement scheme that permits to fully
characterize vortex states in low density atomic condensates, yielding
both direct evidence of persistent currents as well as a parameter-free
determination of the topological charge.

\acknowledgements
This work is supported in part by the U.S. Office of Naval Research
Contract No. 14-91-J1205, by the National Science Foundation Grant
PHY95-07639, by the U.S. Army Research Office and by the
Joint Services Optics Program.

%\bibliography{quasi_new}

\end{document}